\documentclass{acta}
\usepackage{supertabular,lscape,epsfig}
\usepackage{amssymb}
\usepackage{amsmath}
\usepackage{graphicx}
\usepackage{caption}

\SetPages{0}{0}

\SetVol{65}{2015}

\begin{document}

\begin{Titlepage}
\Title{Robust Filtering of Artifacts in Difference Imaging for Rapid Transients Detection}
\Author{J.~~K~l~e~n~c~k~i,~~
{\L}.~~W~y~r~z~y~k~o~w~s~k~i,~~
Z.~~ K~o~s~t~r~z~e~w~a~--~R~u~t~k~o~w~s~k~a~~
and~~ A.~~U~d~a~l~s~k~i}
{Warsaw University Observatory, Al.~Ujazdowskie~4, 00-478 Warszawa, POLAND\\
e-mail:jklencki@astrouw.edu.pl}

\end{Titlepage}

\Abstract{Real-time analysis and classification of observational data
collected within synoptic sky surveys is a huge challenge due to
constant growth of data volumes. Machine learning techniques are often
applied in order to perform this task automatically. The current
bottleneck of transients detection in most surveys is the process of
filtering numerous artifacts from candidate detection. We present a new
method for automated artifact filtering based on hierarchical
unsupervised classifier employing Self-Organizing Maps (SOMs). The
system accepts 97\% of real transients and removes 97.5\% of artifacts
when tested on the OGLE-IV Transient Detection System. The improvement
of the artifacts filtering allowed for single-frame based rapid
detections of transients within OGLE-IV, which now alerts on transient
discoveries in less than 15 minutes from the image acquisition.}

\section{Introduction}

Over the last years, with large-scale sky surveys in their prime, the
detection of rare supernovae-like transient events has been, to a large
extent,  successfully solved.  Projects such as the Palomar Transient
Factory (PTF, Law \etal 2009), the Catalina Real-Time Transient
Survey (CRTS, Drake \etal 2009), PanSTARRS (Kaiser \etal
2010), the All-Sky Automated Survey for SuperNovae (ASAS-SN, \eg Shappee
{\it et al.} 2014), the Optical Gravitational Lensing Experiment (OGLE,
Wyrzykowski {\etal 2014a, Udalski \etal 2015), Gaia Science
Alerts program (Wyrzykowski \etal 2014b) and others are reporting
numerous transient discoveries on a daily basis.

The remaining challenge, however, is to bring the transient  detection
process as close to real-time as possible. The quicker we discover an
ongoing transient event, the sooner we are able to publish an alert,
prompting extra follow-up observations. Early discovery of an ongoing
event can also allow obtaining valuable spectroscopic observations (\eg
PESSTO survey, Smartt \etal 2015). Ultimately, thanks to the
follow-up started early, one can obtain far better data coverage than
that one would achieve when identifying the event only at, or after, its
maximum brightness, which is, in turn, leading to better scientific
results.  For example, in the planetary microlensing detection, it is
crucial to detect and intensively observe a very short-lived deviation
from the standard model caused by a low-mass companion to the lens (\eg
Udalski \etal 2005, Beaulieu \etal 2006, Skowron {\it et
al.} 2015) in order to properly characterize the properties of the
planet. 

When working with supernovae (SNe) -- which are usually the key focus of
most transient surveys thanks to their cosmological applications (\eg
Sullivan \etal 2011, Campbell \etal 2013 using SN Ia as
"standardizable" candles following Phillips 1993) -- good quality
photometric and spectroscopic observations are essential for accurate
standardization of SNe events, as well as for better understanding of
the relations between characteristic properties of each event (\ie
supernova's itself, its environment, etc.) and observed features in its
light curve and spectra (\eg  Childress \etal 2013a, 2013b,
Galbany \etal 2014). Although thousands of supernovae events have
been found up to date, only a small part of them have good enough data
coverage from the very early days since explosion, sufficient to be
studied effectively.  

Other motivation for real-time supernovae detection is that we lack data
points from the earliest stages of the events and these are the
observations that carry information about the actual physics of
supernovae explosion (\eg early-time spectrum revealing W-R-like wind
signatures, Gal-Yam \etal 2014). Being able to obtain good
quality photometry from before the maximum of brightness would allow for
better verification of theoretical models of the most common SNe types,
as well as possibly help understand the physics behind rarely observed
exotic phenomena, such as, for example, super-luminous supernovae (\eg
Quimby \etal 2011, Nicholl \etal 2015a, 2015b)

While the time scale of 'real-time' naturally depends on the type of
transients we are looking for -- varying from seconds in Gamma-Ray
Bursts regime to days for long microlensing events caused by  black
holes -- the constantly growing huge volumes of data are common for all
kinds of sky surveys.  With even bigger observational projects planned
for the future (\eg LSST, Ivezi\'{c} \etal 2008), it is clear that
automation of data processing and analysis is inevitable.  Current
surveys' pipelines are already equipped with algorithms to perform
observations, data reductions and calibrations  automatically.  Also the
search for any changes in the brightness of observed objects, often
carried out by time-domain surveys based on differential imaging
technique, is done by the computers, resulting in numerous candidates
for transient events.  The remaining challenge of today's pipelines is
to automatically perform the next stage, which is the actual filtering
and classification of the vast volumes of the observational data. 

Given the view of real-time transient detection, the advantages of
computational (rather than human-based) technique are quite substantial:
\begin{itemize}

\item properly trained algorithms are much faster than human
astronomer(s) analyzing individual data case by case. This allows
triggering almost instantaneous follow-up observations and more
effective usage of limited resources (observational time at telescope
facilities) in order to obtain better data coverage;

\item computational classification is deterministic, which also allows
calibration of uncertainty. It appears that even experienced astronomers
can have various opinions concerning same detection, which makes
standardization uneasy (see Bloom \& Richards, 2011)

\end{itemize}

The task of the automation itself, however, is not straightforward. The
most promising approach is to implement machine-learning techniques.
Many papers have been published recently on the subject, presenting
automatic classification systems involved in different sky surveys.
Wright \etal (2015) apply various machine-learning techniques for
transient discovery in Pan-STARRS1 difference imaging, while Goldstein
\etal (2015) searches for transients in the Dark Energy Survey. Eyer \&
Blake (2005) train the unsupervised Bayesian classifier of variable
stars from the ASAS survey. S\`{a}nchez \etal (2010) employ
$k$-means clustering to classify galactic spectra from the SDSS survey.
Dubath \etal (2011) show how Random Forests can be applied in
classification of variable stars from the Hipparcos mission. Bloom {\it
et al.} (2012), by also implementing the Random Forest algorithm,
develop the automated system for data analysis and preliminary
classification in the PTF survey. Blagorodnova \etal (2014) take
a Bayesian approach to classify transient events discovered by the Gaia
satellite based on their low-resolution spectra.  An excellent overview
of the data mining and machine learning techniques in time-domain
astronomy is presented by Bloom \& Richards (2011). A more brief
discussion of real-time transient search specifically can be found in
Mahabal \etal (2012).

A specific type of classification problems is artifacts filtering. 
Finding candidates for transient events is very often overwhelmed by a
huge number of unwanted artifacts, caused by cosmic rays, satellites'
crossings, problems in reduction of images, misalignments of the images,
CCD defects and others.  Dealing with artifacts is extremely
time-consuming as most of the work has to be done manually. Thus, the
stage of selecting valuable candidates for transients is the current
bottleneck of any transient survey and performing it automatically is an
essential step towards real-time detections.  In this paper we present a
new method for robust, automatic artifact filtering and a transients
detection system based on a hierarchical classifier of Self-Organizing
Maps (SOMs). The most obvious advantage of SOMs is that they do not
require preparation of any pre-classified and labeled training set which
helps avoid potential biases introduced via the training set in
supervised methods. The method has been applied to improve the detection
purity and speed in the Optical Gravitational Lensing Experiment (OGLE)
survey for transients (Wyrzykowski \etal 2014a) in its new, rapid
generation.

The paper is organized as follows. In Section 2 we describe how the
transients are searched in the OGLE-IV survey, where the data we use
here come from. Section 3 shows how the artifact classifier is
constructed. Section 4 describes the application of the new classifier
to the new rapid transient search in OGLE-IV. We conclude in Section 5.

\section{Transients Detection in OGLE-IV}

The Optical Gravitational Lensing Experiment has been started in 1992
(first phase: OGLE-I) as one of the first surveys dedicated to
gravitational microlensing events. The current  phase of the project
(OGLE-IV, since 2010) is conducted on a 1.3 meter Warsaw Telescope
located at the Las Campanas Observatory, Chile, operated by the Carnegie
Institution for Science.  The telescope is equipped with a large field
CCD mosaic camera with 32 CCD detectors (2x $\times$ 4k each) -- one of
the largest CCD mosaics worldwide. It covers the entire field of view of
the Warsaw telescope  (1.4 square degrees; scale 0\zdot\arcs26 /pixel).
The OGLE-IV project carries out observations in two filters of
Johnson/Cousins photometry system: the $I$- and $V$-band, with the
majority of observations being done using the former.  More technical
details of the survey can be found in  Udalski \etal (2015).

OGLE-IV observations cover the Galactic bulge, disk and the Magellanic
Clouds and their surroundings.  The extragalactic transients are being
search for in the region covering about 650 square degrees toward the
Magellanic Clouds System (MCS). Some of those regions have been
regularly observed since 2010, which resulted in a huge database of
objects with variability monitored over years.  The typical cadence of
MCS observations is from two to five days, depending on the time in the
season (typically from August until April), resulting in few hundreds of
frames collected per field.

The area around Magellanic Clouds is observed primarily in the $I$-band
and with exposure of 150 seconds.  Read-out time for 32-CCD-chip mosaic
is about 20 seconds (Udalski et al. 2015). In the next minute after data
acquisition the reduced (bias-subtracted and flat-fielded) image is
ready for further processing.

Since the primary motivation for creating the OGLE project was the
search for microlensing events, which occur in the most dense stellar
regions, the OGLE have mastered the difference imaging technique to
achieve the best results in those challenging sky regions. The
Difference Imaging Analysis (DIA) method tuned to the OGLE data is based
on the Wo{\'z}niak (2000) implementation of Alard \& Lupton (1998)
algorithm.  The reference images for the DIA method are obtained by
stacking several high-quality images obtained under perfect seeing
conditions (less than 1\arcs). 

The data processing pipeline reduces incoming images on-the-fly and
after about three minutes the subtracted image is produced along with
detections of new sources, their positions and preliminary magnitudes. 
The match to Gaussian profile of each brightening must exceed 0.7, with
1.0 interpreted as a perfect match, as we are looking for point sources.
Subtractions fulfilling those conditions are accepted as candidates,  
cross-matched with OGLE's database of known stellar-like objects and
divided into two channels: \textit{old sources} (\ie brightenings of
previously known objects) and \textit{new sources} (\ie brightenings
appearing as new objects).  Because most of the galaxies are not in the
OGLE's stellar-like objects database the investigation of candidates
from the new source channel is ideal for the search for exogalactic
transients such as supernovae. Brightenings detection from the old
source channel, on the other hand, are mostly caused by variable stars
and cataclysmic variables.

The search for transients in years 2012--2014 (Wyrzykowski \etal
2014a) required two subsequent detections of a candidate transient to
ensure high purity of discoveries.  The selected candidates (typically
100--300 each night) were visually inspected and promising transients
were picked based on contextual information (\eg galaxy nearby) and
archival data (previous variability or non-detections). 

\section{Artifact Filtering Classifier}

As can be expected, the vast majority of transient candidates from data
reduction pipeline are unwanted artifacts. The most common are four
types of artifacts: (a) satellites' crossings, (b) cosmic rays, (c)
image subtraction defects and (d) ''yin-yang'' shapes caused by stellar
objects with high proper motion.  Fig.~1 shows examples of artifacts
as seen in the difference images produced by the OGLE pipeline. 
Avoiding the most common types of artifacts, types (a) and (b), can be
easily achieved by the requirement for repeating observations of the
same field. However, such method stands against the idea of bringing
transient detection as close to real-time as possible, as in OGLE's case
the detectability time frame is then limited by our cadence of 2--5 days.
 This is a trade-off between the number of repeated observations versus
the sky area covered.  In order to allow for very rapid discovery of
transients with the current OGLE observing strategy we aimed at
filtering the artifacts based on a single OGLE observation. 

\begin{figure}[htb]
\includegraphics[width=3cm]{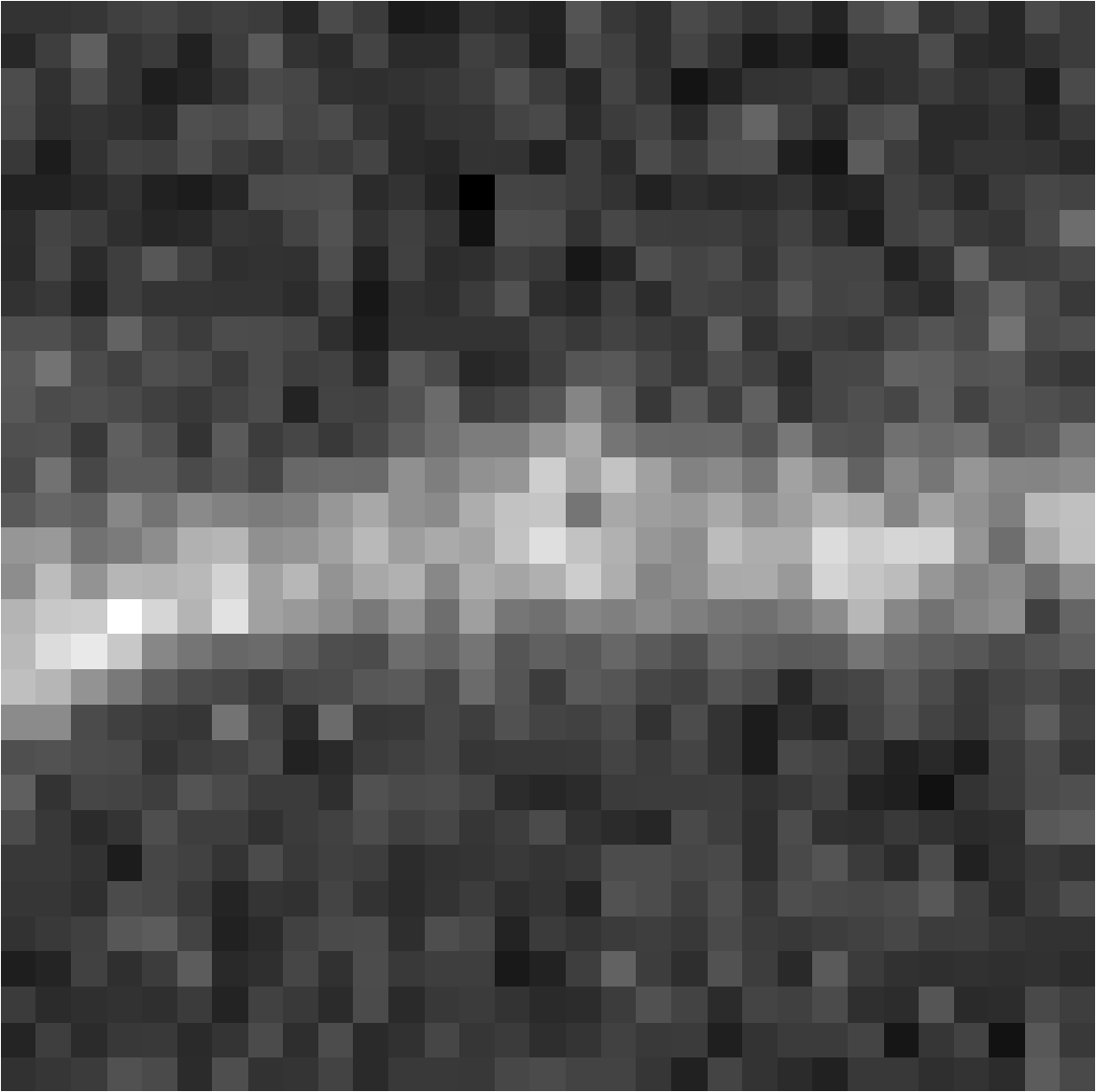}\hskip2mm
\includegraphics[width=3cm]{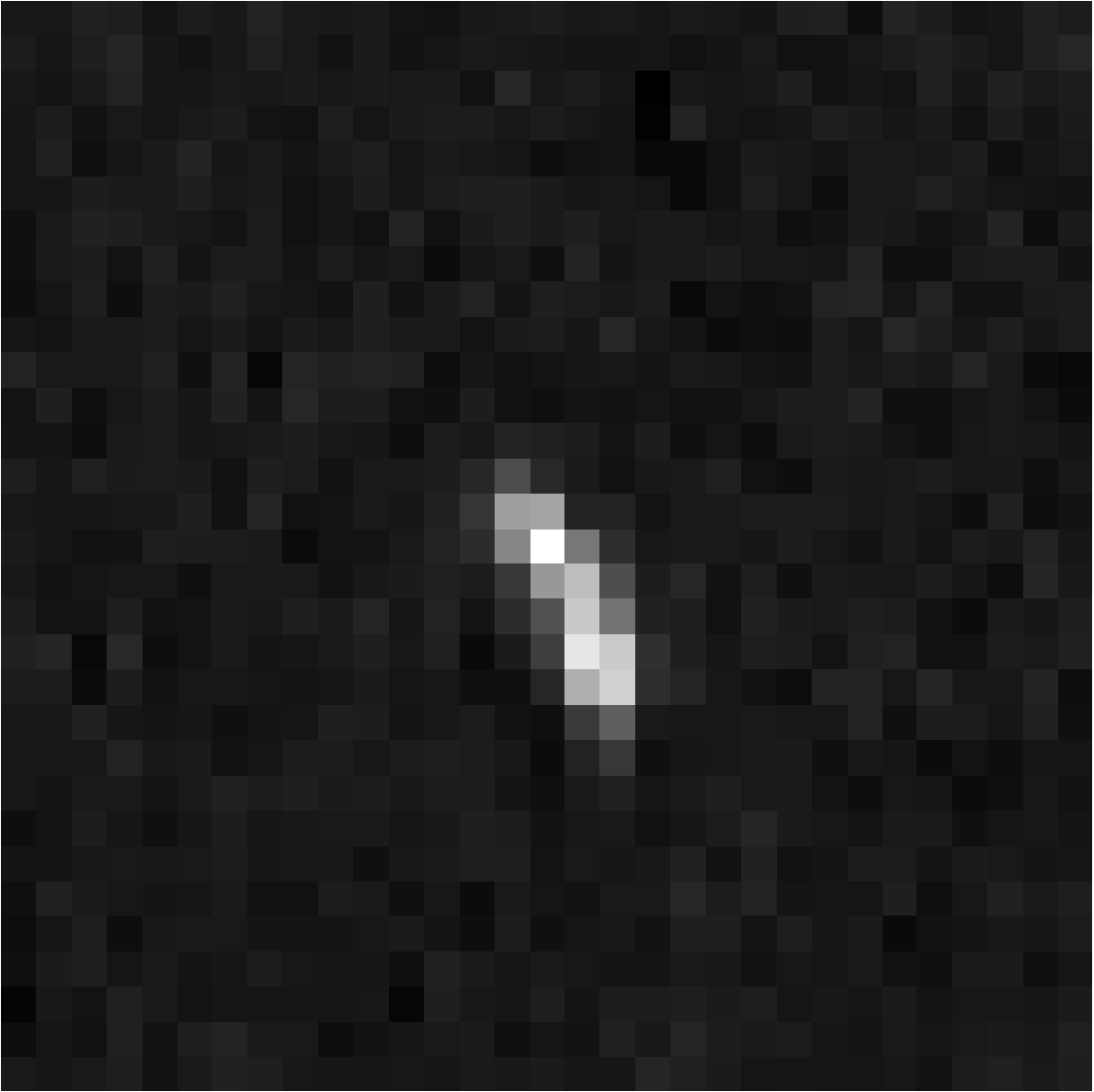}\hskip2mm
\includegraphics[width=3cm]{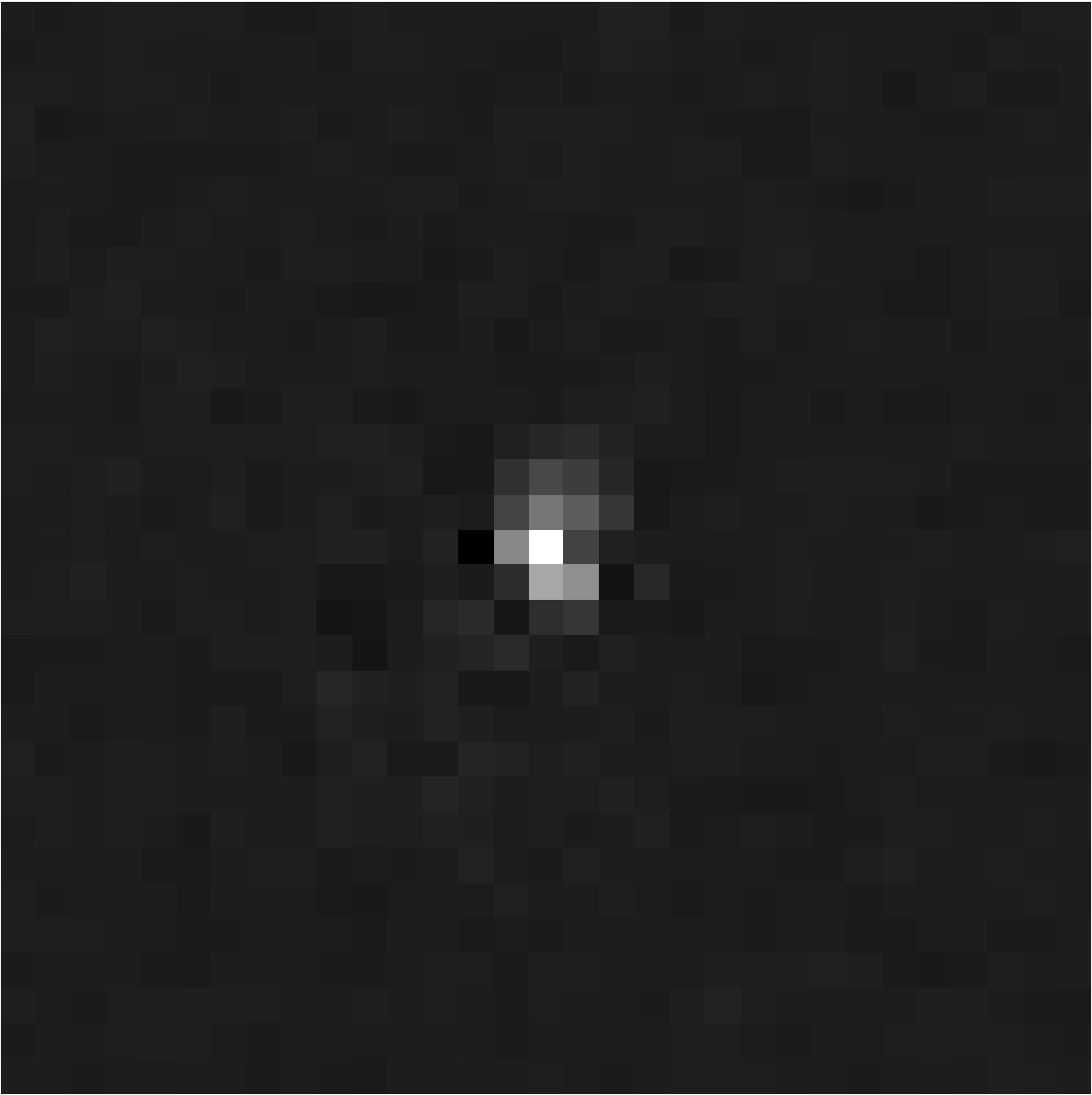}\hskip2mm
\includegraphics[width=3cm]{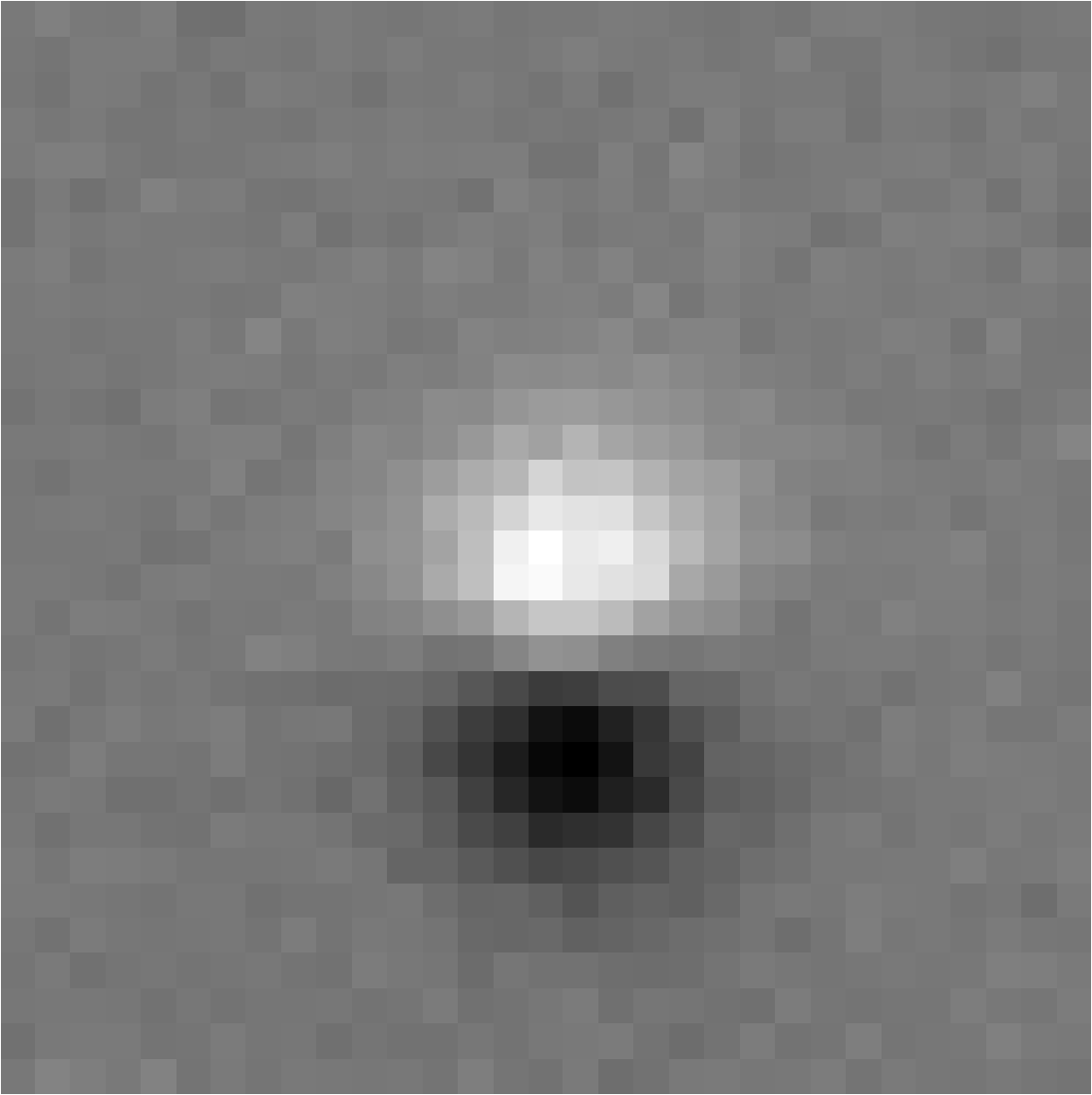}
\vskip7pt

\FigCap{Most common types of artifacts from the OGLE transient pipeline
presented as $31\times31$ pixel cutouts from difference images. {\it
Left panel} shows satellite's crossing artifact and {\it two next
panels} -- cosmic ray and subtraction failure (too faint) artifacts. The
{\it most right} panel presents high proper motion stellar object
artifact.}

\end{figure}

\subsection{The Tool: Self-Organizing Map}
Self-Organizing Map (SOM, Kohonen 1982), is an example of an
unsupervised artificial neural network, which has a huge potential in
astronomical data analysis problems (see Wyrzykowski \& Belokurov,
2008). The ability to organize itself helps when dealing with classes of
objects with not well defined boundaries.  The network is usually
arranged into a two-dimensional array of nodes with \textit{weight
vectors} $\mathbf{w}$ associated to them.  Each analyzed datum, in our
case a 31x31 pixel cutout image of brightening, is represented in form
of a numeric vector $\mathbf{t}$ (of same length as weight vectors) and
mapped onto node with weight vector $\mathbf{w_0}$ most similar to it --
the winner node. The usual criteria for similarity is the least
Euclidian distance between vectors,  \ie the following  condition is
fulfilled:

\begin{equation}
 \forall_ {\mathbf{w} \in SOM} \; \; \; |\mathbf{w}-\mathbf{t}| \geqslant |\mathbf{w_0}-\mathbf{t}| 
\end{equation}

The SOM organizes itself during an unsupervised learning process. The
initial values of weight vectors can be set according to some pattern,
for instance linearly growing. However,  for a sufficiently long
training loop the initial conditions do not affect the final form of SOM
in any noticeable way. Then, for each training vector $\mathbf{t}$
(chosen either randomly or one by one from the training dataset) the
winner node $\mathbf{w_0}$ is found. The actual learning process takes
place now, when the winner $\mathbf{w_0}$ and all its neighboring nodes
are adjusted with a learning rate  $\mathbf{\alpha (n)}$ in order to
become more similar to the training vector $\mathbf{t}$. For the step
number $n$ of the training loop the learning process will be as follows:

\begin{equation}
 \mathbf{w_{new}} = \mathbf{w_{old}} + \alpha(n) (\mathbf{t} - \mathbf{w_{old}})	\; \; 
\end{equation}

The learning rate $\mathbf{\alpha (n)}$ decreases from 1 to 0 during the
whole training process, typically in a linear way. 

\subsection{Data Representation}
The key element of a SOM-based classifier is the way of representing
each analyzed datum in form of a numeric vector. It should contain all
the  information needed for a successful classification and in the same
time the vector should not be too long (longer vector means longer
calculation time).  In this case for each transient candidate we first
produce a 31x31 pixel cutout from the difference image centered on the
brightening. The image values are normalized in such a way, that the
brightest pixel  always has the value $1.0$ and the mean background
level corresponds to $0.5$ -- this means that we only analyze the
profile (shape) of the brightening, without considering  the absolute
value of brightening at this stage. We also perform the \textit{angle
normalization} of cutouts, that is rotate each of them by an angle of
$0^{\circ}$, $90^{\circ}$, $180^{\circ}$ or $270^{\circ}$ in such a way,
that the brightest half-part of the image (\ie having biggest sum of
pixel values) is always the top one. This way we reduce by half the
number of somewhat artificial types of brightenings appearing due to
their random orientation. Normalized cutouts are then  represented in
form of numeric vectors of length $1061$: the first $961 = 31^2$ fields
are simply the values of the cutout's pixels and the last $100$ fields
correspond to heights of  the bins of a normalized histogram of the
image. This representation, similar to that adopted by Wright {\it et
al.} (2015) for transient detection in the Pan-STARRS1 survey, differs
from most of previous work in that it works directly on the pixel's
values rather then on specific features of the image.

\subsection{Hierarchical Classifier} 

We trained a triple-stage SOM classifier for transients detection from
OGLE. The first stage (\textit{Anti-cosmic SOM}) is responsible for
ruling out the majority of the artifacts of most numerous types:
satellites' crossings and cosmic rays.  At the second stage
(\textit{Anti-yinyang SOM}) the classifier eliminates most of the
''yin-yang''-like objects and the remaining cosmic rays.  The final
stage map (\textit{Anti-faint SOM}) is aimed at dealing with those types
of artifacts that the second stage SOM failed to exclude -- which are
mostly image subtraction defects appearing as brightening too faint to
be caused by a real transient event. Each stage's SOM has a form of an
independent two-dimensional array of five nodes width and seven nodes
height. 

\vskip 8pt
{\it The first stage: Anti-Cosmic SOM}
\vskip 8pt

We prepared a set of input data based on difference imaging produced by
the OGLE data reduction pipeline.  All new objects (not matched to known
reference objects) brighter than about 19.5 mag were gathered from the
period of a couple of months in early 2014. The training data set
comprised of 7000 detections found based on a single frame observation,
hence consisting primarily of cosmic rays and satellites' crossings. 
The training loop had 50\ 000 steps with learning parameter set as a
constant value $\alpha = 0.2$. The radius of nodes treated as neighbors
was set to three. The trained map is presented in Fig.~2a.

\begin{figure}[htb]
\includegraphics[width=5cm]{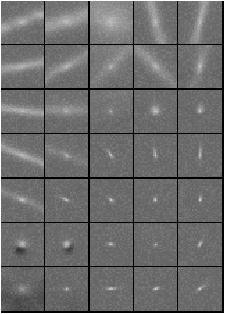}\hskip5mm
\includegraphics[width=5cm]{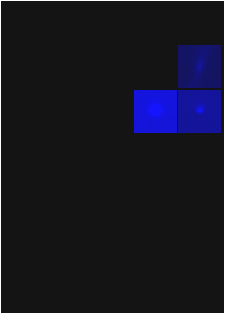}\hskip5mm
\includegraphics[width=1.355cm]{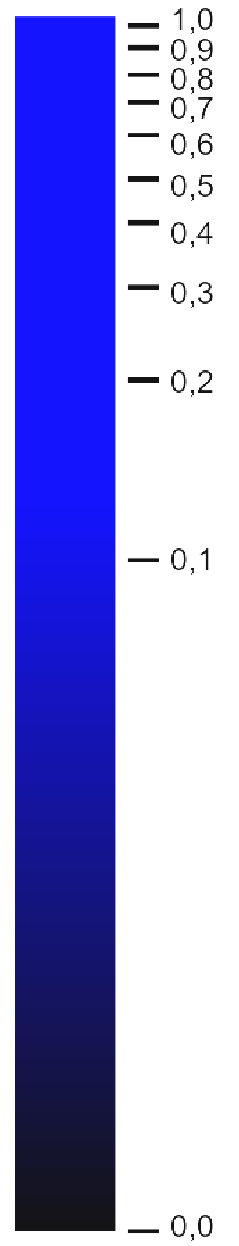}
\vskip7pt
\FigCap{Anti-cosmic SOM -- map trained for the first stage of artifacts filtering
classifier. {\it Left panel} shows visualization of the nodes of the trained SOM.
{\it Right panel} presents map of relative purity of nodes based on the test set.} 
\end{figure}

When using SOM as an artifact filtering tool one has to choose which
nodes form the {\it acceptance region} of the verification system,
\ie transient candidates classified into those nodes are verified as
positive and accepted, whereas the rest is treated as artifacts and
ruled out. In order to determine the best possible acceptance region we
decided, as a rule, to prioritize nodes with highest ''purity'' $P$
(Bloom \etal 2012) defined as:

\begin{equation}
 P = \frac{R_{acc}(real)}{R_{acc}(real) + R_{acc}(artifact)} \; \;,
\end{equation}

The rate of accepting certain ''type'' as ''real'' or ''artifact'' -- can be expressed as:

\begin{equation}
R_{acc}(type) = R(type) \cdot P(acceptance \; | \; type) \; \;,
\end{equation}

which is the rate of occurrence of the type in the analyzed data
population $R(type)$ multiplied by the probability of it being accepted
by the node.

In order to calculate the purity of each node we performed a test
classification on a dataset of more than 13\ 200 images, which contained
89 images of known good transients (chosen manually from transients
selected during the OGLE Transient Search). The probabilities
$P(acceptance \; | \; type)$ for each node were estimated as:

\begin{equation}
P(acceptance \; | \; type) = \frac{\# \;  accepted \; candidates \; of
\; the \; type}{\# \;  all \; candidates \; of \; the \; type \; in \;
the \; dataset} \; \;.
\end{equation}

Based on the test classification we obtained a map of relative nodes'
purity, normalized to range [0,1] (see Fig.~2b). The map clearly shows
that known real transients landed in only three nodes (purity larger
than 0). Those three nodes accepted altogether all 89 real transient
detections and 599 of nearly 13\ 200 artifacts, which corresponds to the
false positive ratio (FPR) of  $FPR_{I \; stage} \approx 0.045$ and the
true positive ratio (TPR) being $TPR_{I \; stage} = 1.0$. Naturally,
these nodes formed the acceptance region for the first stage of the
classifier.

\vskip 8pt
{\it The Second Stage: Anti-Yinyang SOM}
\vskip 8pt

The second stage SOM has been trained on a different training dataset
than in the first stage. It contained about 12\ 000 images of detections
from the OGLE pipeline which were selected requiring at least two
independent observations at the same location on two subsequent frames,
again with 19.5 mag threshold for both of the detections.  The brighter
of the two was chosen for the training set. Thanks to that selection
most of the artifacts caused by cosmic rays and satellites' crossings
were not present in that dataset. The training loop had 100\ 000 steps
with learning parameter set as a constant value $\alpha = 0.2$. The
radius of nodes treated as neighbors was set to three. The trained map
is presented in Fig.~3a.

\begin{figure}[htb]
\includegraphics[width=5cm]{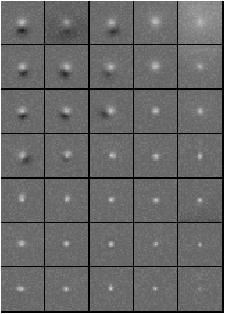}\hskip5mm
\includegraphics[width=5cm]{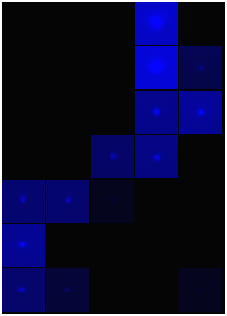}\hskip5mm
\includegraphics[width=1.355cm]{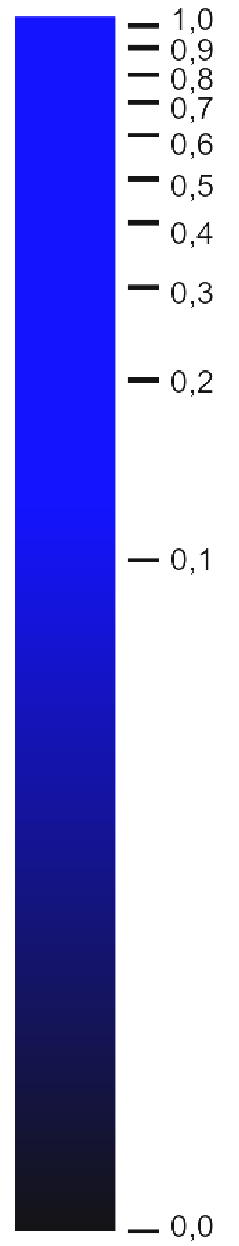}
\vskip7pt
\FigCap{Anti-yinyang SOM -- map trained for the second stage of artifacts filtering
classifier. {\it Left panel} shows visualization of the nodes of the trained SOM.
{\it Right panel} presents map of relative purity of nodes based on the test set.} 
\end{figure}

The acceptance region was determined based on a test classification of
over 12\ 400 detections, requiring their presence on the same location in
at least two subsequent frames.  It contained transient events detected
and confirmed by long term monitoring by the OGLE in years 2012--2014
(total of 327 real transients).  Based on the classification we obtained
a map of normalized relative nodes' purity, following the procedure
described above -- see Fig.~3b.  At this stage there were multiple nodes
with positive purity. We included all of them to the acceptance region,
resulting with all 327 real transient detection as well as 6859
artifacts being  accepted, which corresponded to the false positive
ratio of $FPR_{II \; stage} \approx 0.57$ and true positive ratio of
$TPR_{II \; stage} = 1.0$.  It shows that it is much more  difficult to
obtain pure classifier performance at the second stage in comparison to
the first stage ($FPR_{I \; stage} \approx 0.045$ and $TPR_{I \; stage}
= 1.0$), as  the most obvious and easy to recognize artifacts were
already filtered out at the first stage. 

\vskip 8pt
{\it The Third Stage: Anti-Faint SOM}
\vskip 8pt

To improve the performance of the non-cosmic artifacts elimination from
the second stage we trained the third, final stage SOM aimed at
recognizing those types of defects the second stage SOM failed to
distinguish from real transient detections.  The training set was thus
composed of the 7186 candidates that were accepted at the second
stage's test classification:  327 confirmed transients and 6859
artifacts. The training loop had 100\ 000 steps with learning parameter
set as a constant value $\alpha = 0.2$. The radius of nodes treated as
neighbors was set to three. The trained map is presented in Fig.~4a.

\begin{figure}[htb]
\includegraphics[width=5cm]{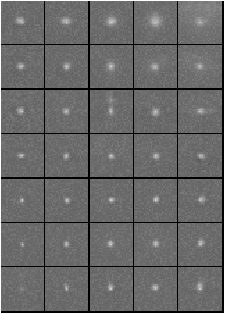}\hskip5mm
\includegraphics[width=5cm]{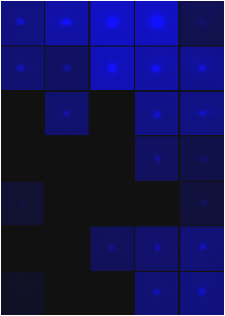}\hskip5mm
\includegraphics[width=1.355cm]{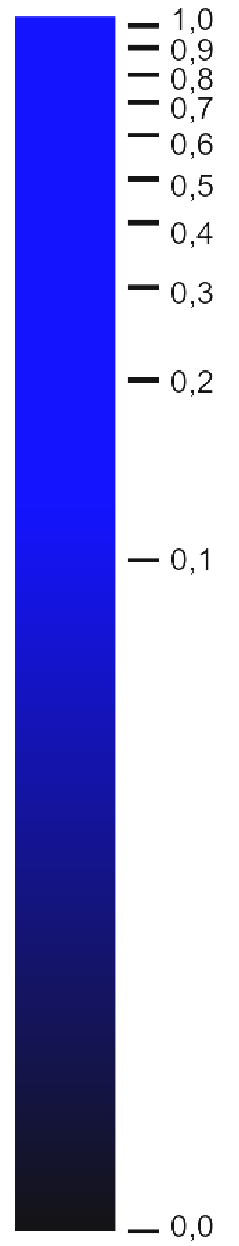}
\vskip7pt
\FigCap{Anti-faint SOM -- map trained for the third stage of artifacts
filtering classifier. The training dataset consisted of 7186 images of
OGLE detections that were passed by the second stage SOM. {\it Left
panel} shows visualization of the nodes of the trained SOM. {\it Right
panel} presents map of relative purity of nodes based on the test set.} 
\end{figure}

The acceptance region was determined based on a test classification of
same database of 7186 candidates. A map of normalized relative nodes'
purity was drawn following the procedure  described in Section 3.4 --
see Fig.~4b. At the final stage the choice of acceptance region is more
difficult than at previous stages as most of the nodes express positive
purity.  The more nodes we included into the acceptance region, the
lower the false negative ratio (FNR = 1 -- TPR), however, the false
positive ratio increases alongside -- something we would very much like
to avoid. This fact is illustrated by the receiver operating
characteristic (ROC) curve, which plots FPR with respect to FNR for
different  sizes of the acceptance region (see Fig.~5). 

\begin{figure}[htb]
\centerline{\includegraphics[width=10.5cm]{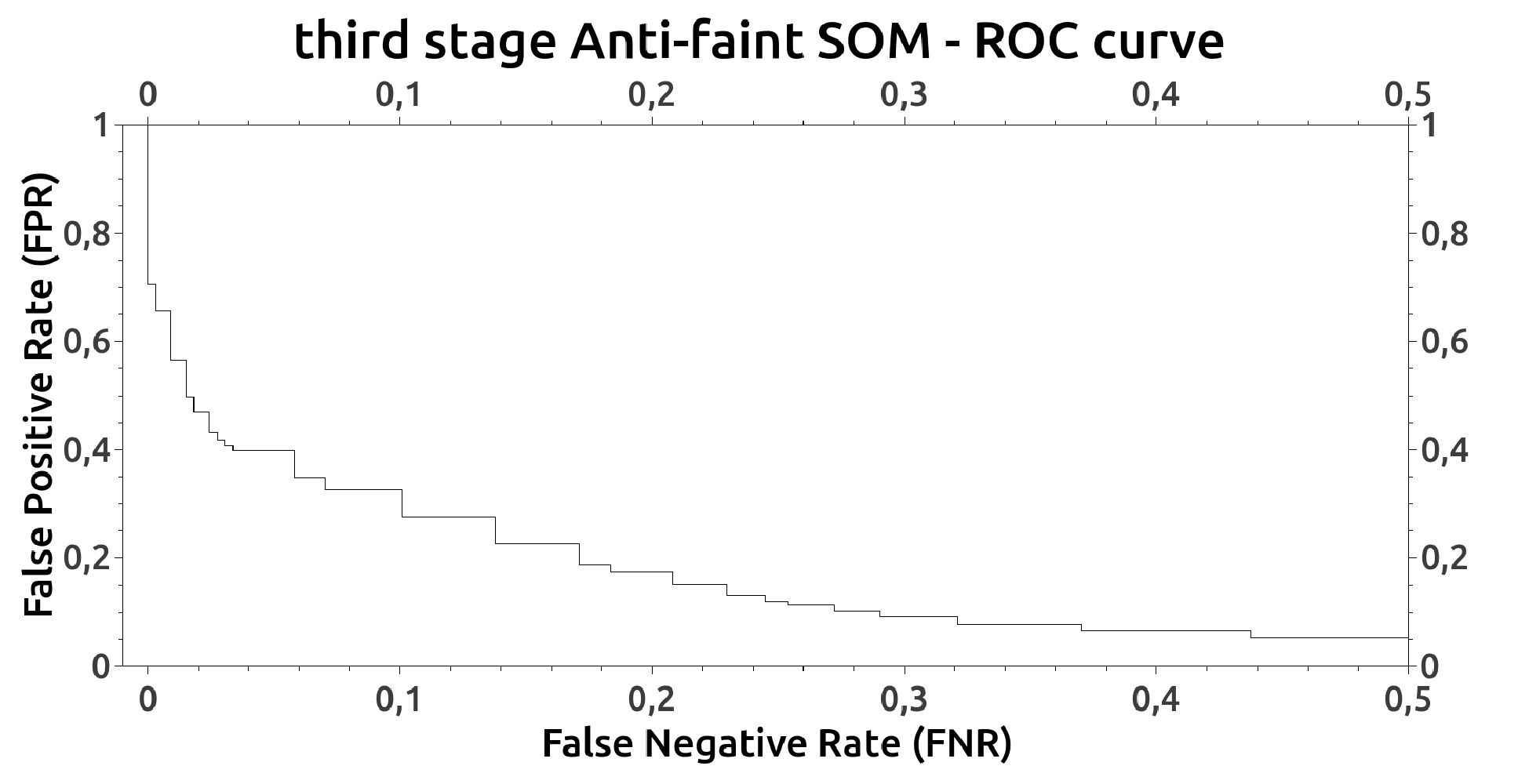}}
\vskip7pt
\FigCap{Receiver operating characteristic (ROC) curve for the third-stage SOM.}
\end{figure}

\subsection{Triple-Stage Classifier's Performance}

In order to decide on the number of nodes forming the acceptance region
at the third stage we investigated the behavior of the fully operating,
triple-stage classifier. We performed test classification of over 13\
000 candidate detections from the OGLE pipeline gathered without the
requirement of having at least two independent observations, \ie based
solely on a single observation. The acceptance region at the first and
second stage were set as discussed above. By varying the size of the
third stage acceptance region we obtained different FPR to FNR ratios
for the entire triple-stage classifier, as presented in form of the ROC
curve in Fig.~6.

\begin{figure}[htb]
\centerline{\includegraphics[width=10.5cm]{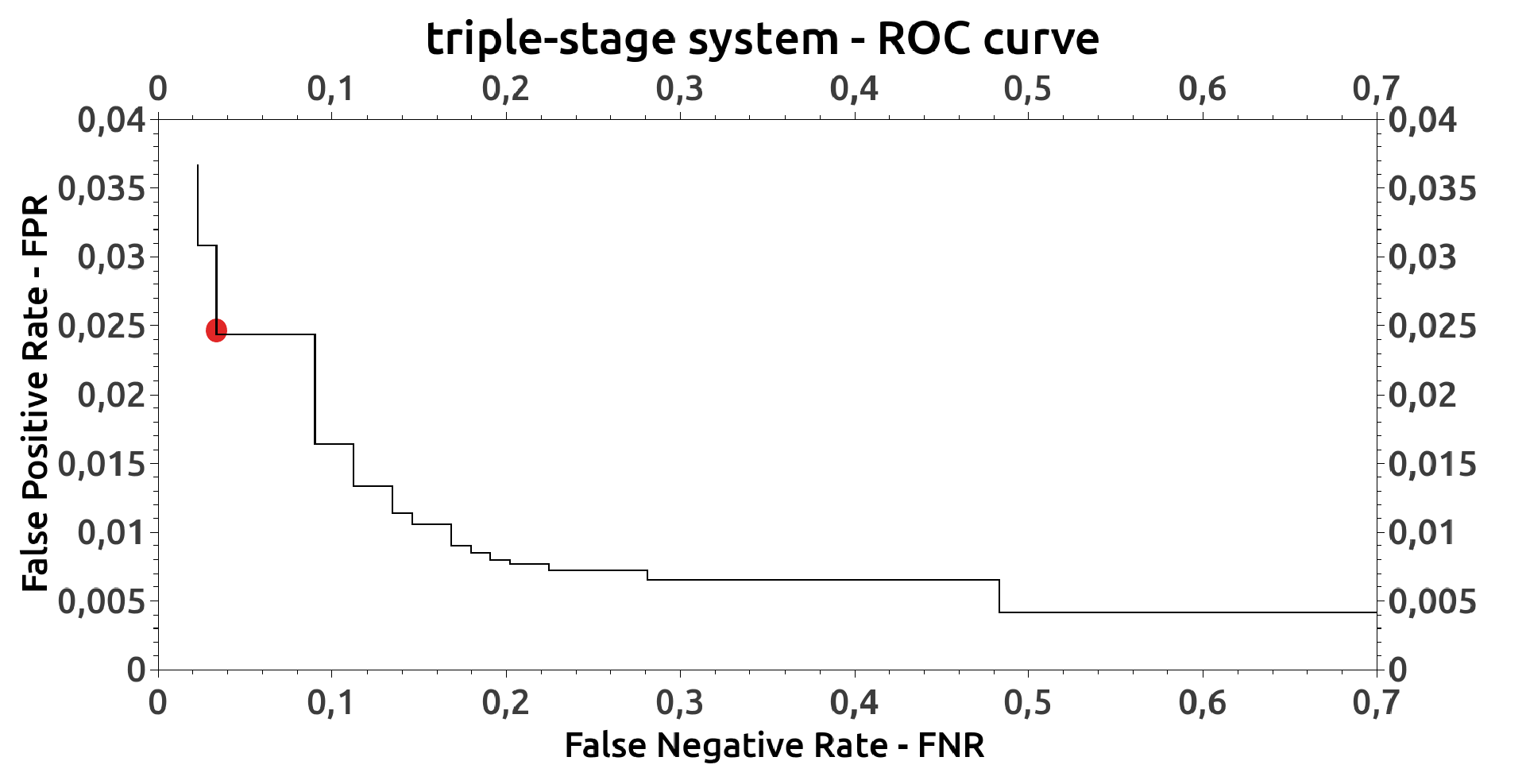}}
\vskip7pt
\FigCap{Receiver operating characteristic (ROC) curve for the entire
triple-stage artifact filtering classifier based on result of a test
classification of 13\ 000 candidate detections from the OGLE pipeline,
generated without the requirement of having at least two independent
observations. For the chosen threshold (red dot) the system accepts
$\approx 97\%$ of real transient detections and excludes $\approx97.5\%$
of artifacts from the data.}
\end{figure}

For an exemplary and satisfactory threshold (marked on the ROC curve
with a red dot) the system accepts $\approx 97 \%$ of real transient
detections and excludes $\approx 97.5 \%$ of artifacts from the data.
The purity of the candidates database (here the percentage of real
transient detections) has increased to the final purity of $\approx
26\%$ from the initial value of 0.7\%. This is a huge improvement, as
the remaining number of daily candidates can be  easily inspected and
verified manually in close to real-time time-frame.  It is worth noting
that the classifier relied solely on the difference images and did not
employ any light curve nor contextual information around the transient
candidate.

\section{Application: Rapid OGLE Transients Detection System}

OGLE-IV has been running a search for transients objects, OGLE
Transients Detection System (OTDS, Koz{\l}owski \etal 2013, Wyrzykowski
\etal 2014a) since 2012.  This dedicated survey has discovered so far
about 500 various transients, primarily supernovae and novae.  The
spectroscopic classification and follow-up has been provided primary
thanks to collaboration with the PESSTO group (Smartt et al. 2015) for
about 100 targets, which resulted in findings of rare and interesting
transients  (\eg Inserra \etal 2015, Poznanski \etal 2015, Pastorello
\etal 2015).

The  triple-stage SOM-based classifier has been applied in a new version
of the OGLE Transient Detection System, the OTDS-Rapid. High efficiency
of the artifact filtering allowed for a substantial increase in the
speed of detection of new transients within the OGLE-IV pipeline. The
original OTDS required two subsequent frames to contain a transient, in
order to filter out numerous contaminants and false detections.  This,
however, resulted in belated discoveries of transients, given our
sampling strategy.  The new OTDS-rapid relies solely on a single frame
exposure taken by the OGLE-IV survey and is able to produce transient
discoveries within less than 15 minutes after data acquisition. 

The ODTS-Rapid pipeline analyzes each of the new sources (not linked to
any of the known old OGLE sources during the cross-match). At the
magnitude threshold currently set at 19.5 mag there is from 10 to 50 new
sources detected per chip, meaning 320 to 1600 sources to investigate
with every frame.  For a typical night, there is in total about 25\ 000
new sources to investigate.  After the triple-stage SOM artifact
classifier, there remains about a 100 of candidates.  For those, the
force-photometry is obtained: on known position of the potential
transient we measure photometry on the current and 10 most recent
previous frames.  Only the objects for which there is at least six
previous non-detections are carried over to the next stage. For
transients located at galactic cores we rule out AGN-like object using
WISE colors, applying the criteria of Assef \etal (2010). We also
compute the offset from the nearest galaxy-like object, which were
detected on the OGLE reference images (Kostrzewa-Rutkowska \etal in
prep.). The accuracy of the offset computation is better than
0\zdot\arcs13 (Wyrzykowski \etal 2014a). In a typical night there is
about 10 to 30 candidates selected in a completely autonomous manner.
The webpage:

\begin{center}
{\it http://ogle.astrouw.edu.pl/ogle4/transients/rapid/rapid.html}
\end{center}

\noindent
is updated every 5 minutes and contains the most recent discoveries
carried out by the OTDS-Rapid, which are available for the astronomical
community to select targets for further immediate follow-up. Fig.~7
shows an example discovery of OGLE15hy, which turned out to be a type Ia
supernova. The time from the observation to the moment the object
appeared on the webpage was about 15 minutes. 

\begin{figure}[htb]
\includegraphics[width=3.8cm]{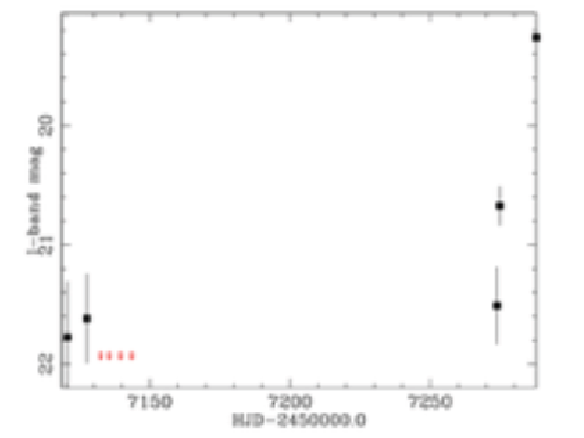}\hskip4mm
\includegraphics[width=8.5cm]{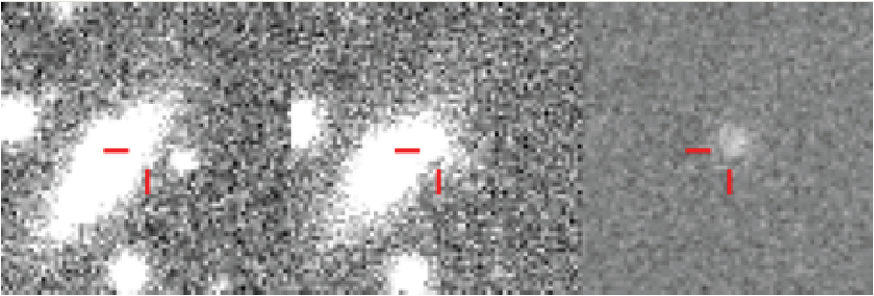}
\vskip7pt
\FigCap{Example rapid detection of OGLE15hy type Ia supernova in about
15 minutes after the observation. {\it Left panel} shows the light curve
at the moment of the discovery with magnitude threshold at $I=19.5$.
Pre-discovery data points were obtained by force-photometry at the
position of the transient. The red arrows indicate non-detections. {\it
Right panel:} reference, current and difference image of the transient.
Size $18\arcs\times18\arcs$.}
\end{figure}

At a conservative threshold of $I=19.5$~mag we are already able to
detect supernovae type Ia at about 2 magnitudes before their maximum
brightness for a typical redshift in our survey of $z = 0.07$.  For a
more ambitious threshold of $I=20.5$~mag (number of artifacts increases
to about 40 every night) we reach a capability of discovering supernovae
type Ia at about --10 to --20 days prior to their maximum.  Rapid
detections help organize the follow-up observations early and
investigate the earliest stages of supernova explosion, providing clues
on supernova origin. 

\section{Conclusions} 

Modern large-scale sky surveys have become very successful in
discovering rare transient events such as supernovae, thanks to
collecting increasingly huge volumes of observational data. Thousands of
candidates for transients are generated every day by repeatedly
analyzing same fields in search for unexpected brightness changes.
Manual verification of so many detections in a short time-frame from the
observation begins to exceed capabilities of observers. Thus, real-time
transient detection -- the ultimate goal of transient surveys -- still
remains a big challenge. The main problem lies in extremely numerous
artifact brightennings among transient candidates,  easy to identify by
a watchful eye of an experienced astronomer but difficult to deal with
automatically.  The usual way of eliminating the most common bogus
detections (randomly appearing cosmic rays and satellites' crossing) is
by requiring the subtraction residual to appear on at least two
independent  sky frames. This approach has a major drawback, however, as
it limits the detectability-frame to the sampling of the survey and
often significantly slows down the detection process. \\

In this work we presented a new method for robust artifact filtering
based on a hierarchical machine-learning classifier composing of
self-organizing maps (SOMs). The triple-stage system  has been trained
on transient candidates from the OGLE Transient Detection System in
order to eliminate the most popular types of artifacts. The classifier
performance has been optimized to a typical population of detections in
the OGLE survey and tested on  a representative dataset of transient
candidates. For a chosen threshold the system accepts $\approx 97 \%$ of
real transient detections and excludes $\approx 97.5 \%$ of artifacts
from the data. This result is satisfactory and sufficient for replacing
the former  requirement for two independent observations and gives birth
to the new Rapid Transient Detection System in the OGLE project.  The
transient detections can now be carried out in less than 15 minutes from
the observation, which is closer to real-time than it has ever been
ever. 

\Acknow{This work was supported by the Polish National Science Center grant
no. 2015/17/B/ST9/03167 to {\L}W. The OGLE project has received funding 
from the Polish National Science Centre grant MAESTRO no. 2014/14/A/ST9/00121 
to AU.}

\end{document}